\documentclass[5p,authoryear]{elsarticle}
\makeatletter 
\def\ps@pprintTitle{%
 \let\@oddhead\@empty
 \let\@evenhead\@empty
 \let\@evenfoot\@oddfoot} 
\makeatother
\usepackage[utf8]{inputenc} 
\usepackage[T1]{fontenc}
\usepackage[fleqn]{amsmath} 
\usepackage{amsthm} 
\usepackage{booktabs} 
\usepackage{multirow} 
\usepackage{amssymb} 
\usepackage{wrapfig}
\usepackage{graphicx}
\usepackage{subcaption}
\usepackage{float}
\usepackage{lipsum}
\usepackage{amsmath}
\usepackage[french]{cleveref} 
\usepackage{natbib}
\setcitestyle{square,numbers}
\begin{document}
\begin{frontmatter}

\title{\textbf{FAULT TOLERANT DIGITAL FILTERS ON FPGA USING HARDWARE REDUNDANCY TECHNIQUES \\ AND DENOISING ECG SIGNALS}}


\author{Dr.Sakthivel SM}
\author{O.Sathvik Reddy}

\address{{School of Electronics and Communication Engineering, Vellore Institute of Technology, Chennai, 60012, India \\

Email id : sakthivel.sm@vit.ac.in, obulreddi.garisathvik2019@vitstudent.ac.in}}

\begin{abstract}

As  more the communications and signal process we use in the today life the more we intend to develop more reliable devices which gives fewer errors due to transient fault, So we use a technique called 5-modular redundancy to generate fewer errors. 5-Modular redundancy is an approach to increasing the reliability of hardware systems constructed from component devices that are subject to failure. In digital signal processing and many other important daily life subjects FIR digital filters are used and  they alike performing multiplications, complex computations and selecting desired frequency for different applications. FIR filters comprise of multipliers, adders, delay units. FIR filters has been chosen because it offers more steadiness and simple execution because of its limited length and no feedback within the circuit.
Fir filter is most important for the signal processing and also used in daily life basis. 
The noise reduction can be performed by denoising of the signals of all the implementations .The proposed FPGA methods use Xilinx Vivado EDA.\cite{trip07}

\end{abstract}


\end{frontmatter}

\section{Introduction}\label{introduction}
The largest problem experienced by  designers of  signal processing, Analog was the capability to break the files over different systems, but however  the results should solidify together in a single result set. This auxiliary issue has remained unsolved for was the fault tolerance both at the processing level and the overall system level  the processing systems. 
N-modular redundancy (NMR) is a technique used in fault-tolerant systems to increase their reliability and availability. It involves replicating a critical component N times and using a voting mechanism to ensure that the system can continue to function correctly even if one or more components fail.\cite{bal47}

In NMR, the replicated components are identical and work in parallel, with each component performing the same function. The outputs of each component are compared to detect any discrepancies, and a voting mechanism is used to determine the correct output. The voting can be done using a majority vote, where the most common output is selected, or a unanimous vote, where all outputs must be the same.
NMR is commonly used in safety-critical systems, such as aerospace, automotive, and nuclear industries. For example, in an aircraft's flight control system, multiple redundant sensors and computers are used to ensure that the plane can continue to fly safely even if one or more than one components fail.\cite{Tar47}
NMR can provide the high levels of reliability and fault tolerance, but it also comes with a cost in terms of the increased complexity, cost, and power consumption.

%
%

Finite Impulse Response(FIR) is a type of digital filter that uses only a finite number of inputs to generate its output, and its impulse response is finite. FIR filter is extensively used in digital signal processing applications. FIR filters are characterized by a set of coefficients, which define the filter's frequency response. The coefficients are typically obtained using a windowing function, such as a Hamming or Kaiser window, and a filter design algorithm, such as the Parks-McClellan algorithm\cite{McC07}.
The basic operation of an FIR filter involves convolving the input signal with the filter coefficients. This process multiplies each sample of the input signal by the filter coefficients and sums the results to produce the filtered output. 

An FIR filter's impulse response is the output of a single input impulse, which is a signal that is zero everywhere except at the time zero, where it has a value of one.
FIR filters have several advantages over other types of filters.\cite{Neh47} They have a linear phase response, which means that they do not introduce phase distortion in the filtered signal. They also have a stable and predictable frequency response, making them ideal for applications such as audio and image processing, where accuracy and the stability are very critical\cite{Mall47}.


\break

\section{PROBLEM STATEMENT}

The problem statement of the Fault tolerant digital filters on FPGA and ECG denoising is as we identify the main outcomes to be generated are the

i)	Less Area is Consumed by the FIR filter architecture

ii)	Less amount of errors to be generated due to fault tolerant devices as we use N-Modular Redundancy Techniques

iii)	The fir filter architecture has to produce ECG signal Denoising using Carry save adders, Multipliers 

In this study, in order to improve the delay of the FIR filter architecture, Vedic Multiplication Process is created to meet the above specifications when developing FIR filters. Additionally, Vedic Multiplier Structures and Carry-Save Adders Structure-Based FIR Filter Design presented, resulting in a smaller area and lowered power consumption. In this  Proposed methodology Vedic multipliers and Carry save adders maintains data accuracy and loss of signal value effectively to generate ECG noise reduction with superior performance metrics. 
.


%
%

\section{5MR Configurations and FIR filter}

\subsection{Conventional 5MR Configuration}

This configuration involves replicating the system or subsystem N times and using a voting mechanism to select the correct output. NMR can detect and correct up to N-1 faults. Here N is considered 5. The conventional 5MR is shown in Fig 1, Here majority voter is used for all outputs.

\begin{figure}[ht]
\includegraphics[width=3.4in]{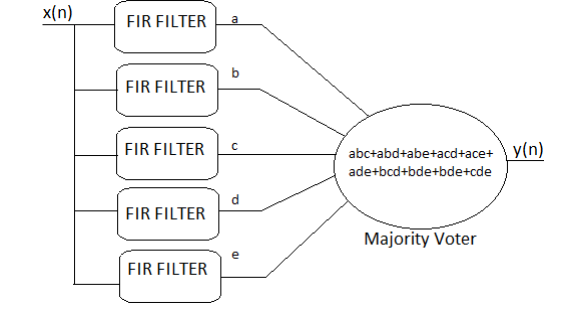}
    \caption[]{Conventional 5MR Configuration} 
\end{figure}

\subsection{5MR as TMR(XOR-MUX) Configuration}

TMR subsystem available in 5MR Systems can be built with different voters. This is the motivation for building different 5MR systems and simulation results by voter type. The TMR subsystem used here is shown in Fig 2. 

\begin{figure}[ht]
	\includegraphics[width=3.5in]{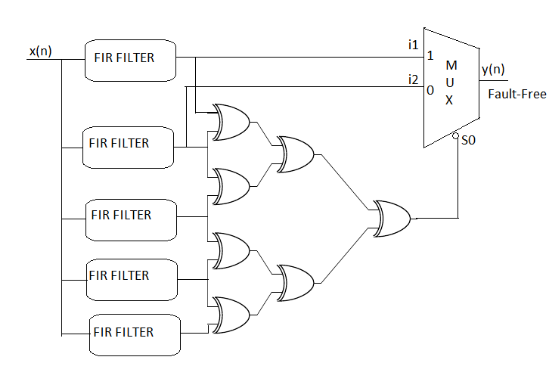}
    \caption[]{5MR as TMR(XOR-MUX)} 
\end{figure}

\subsection{5MR as TMR(XNOR-MUX) Configuration}

Using XNOR with 5MR or TMR provides an additional level of redundancy and improves the reliability of electronic systems. However, it also requires additional hardware, which can increase the overall system cost. As with any design decision, the decision to use XNOR in 5MR or TMR depends on the specific requirements and constraints of the system under design. The Voter of this circuit is similar to the above circuit using XNOR instead of XOR shown in the Fig 3.

\begin{figure}[ht]
	{\includegraphics[width=3.5in]{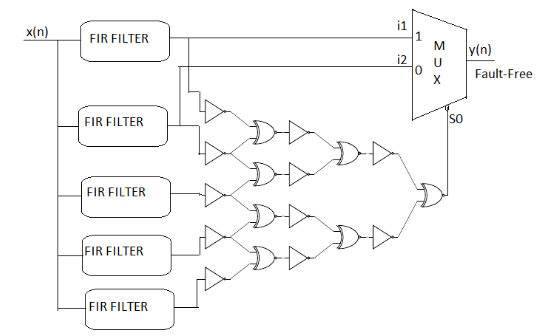}}
	\caption[]{5MR as TMR(XNOR-MUX) Configuration} 
\end{figure}

\subsection{5MR as Cascaded TMR Configuration}

Cascaded TMRs are techniques that combine multiple levels of TMRs to provide even higher levels of reliability and fault tolerance. In cascaded TMR, the output of one TMR plane is fed as the input to another TMR plane. This creates a chain or cascade of TMR circuits, each providing an additional level of error detection and correction. 

TMR affiliated voting department involved 5MR has been replaced by many voters due to its performance comparison. Some selectors may take less layout space and power; few people can afford to consume low power at the expense of area.\cite{Shub47} This type of analysis helps engineers choose the best design with better performance in the system as shown in Fig 4.

\begin{figure}[ht]
	\includegraphics[width=3.5in]{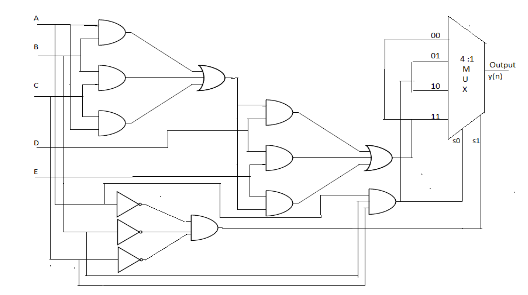}
    \caption[]{5MR as Cascaded TMR configuration} 
\end{figure}

\subsection{5MR with 4 to 1 MUX Configuration}

Using a MUX in this system allows you to select one of the four inputs. This can be used to select the  most likely correct output value. This improves system reliability  by allowing continued operation even if some component fails. This configuration consists of AND gate, OR gate with TMR circuit as well as 4 to 1 multiplexer as shown in Fig 5.

\begin{figure}[ht]
	\includegraphics[width=3.5in]{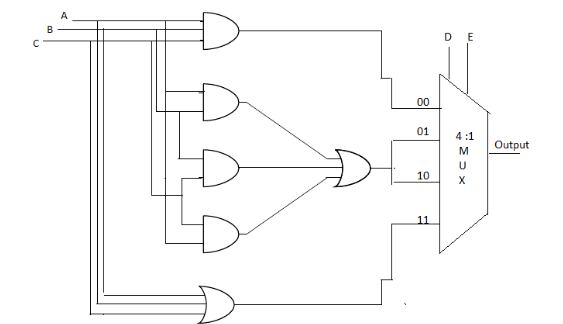}
    \caption[]{5MR with 4 to 1 MUX Configuration} 
\end{figure}

\subsection{Proposed FIR Filter}
In digital signal processing, notably in the study of electrocardiograms (ECGs), FIR filters are frequently utilized. The electrical activity of the heart is represented by an ECG signal, a time-varying signal that has both beneficial and undesirable components. The accuracy of ECG analysis can be improved by using FIR filters to remove unwanted noise and artefacts from the ECG signal.

Using a Vedic multiplier and carry-save adder, FIR filters can be made fast and more effective as it is crucial for real-time signal processing. The carry-save adder minimises the number of additions needed to complete the accumulation, and the Vedic multiplier reduces the amount of arithmetic operations needed to perform the multiplication.\cite{suma47}

Overall, the performance and effectiveness of FIR filters in a variety of applications, including digital signal processing, communications, and image processing, can be enhanced by the employment of specific hardware architectures, such as the Vedic multiplier and carry-save adder,\cite{Sesh47} in this manner the Vedic multiplier, Carry Save Adder and RAM used are of 16-bit as show in Fig 6.\cite{Bak47}.

\begin{figure}[ht]
	\includegraphics[width=3.5in]{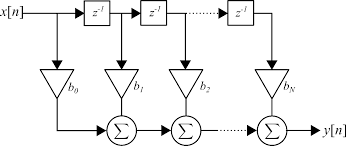}
    \caption[]{FIR Filter} 
\end{figure}

\subsection{Vedic Multiplier}
The 4x4 Vedic multiplier in binary form is implemented using verilog code to reduce delay. 4x4 multiplier designed with 9 full adders and a special 4-bit adder.\cite{Swa47} Architecture of Vedic Multiplier is shown Fig 7  .\cite{Park47}

\begin{figure}[ht]
	\includegraphics[width=3.5in]{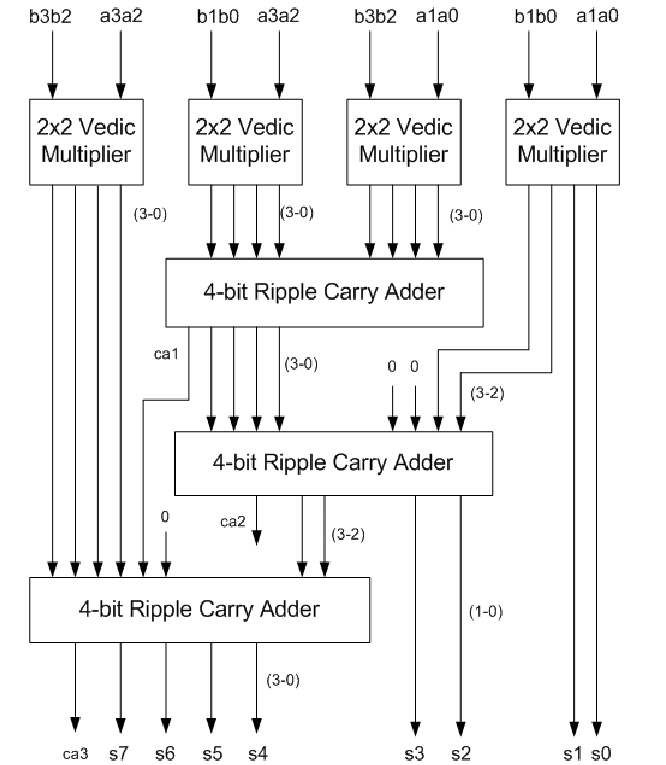}
    \caption[]{Vedic Multiplier} 
\end{figure}

\subsection{Carry save adder}
A CSA produces two partial results, a sum and a carry, by grouping the input integers into bits, then adding each bunch. The amount is then given to the adder's subsequent step, while the carry is kept for use in the future.

The partial sums and saved carries are combined with the total sum in the following step to get the outcome. For bigger numbers, this procedure can be performed numerous times as shown in Fig 8\cite{Sha47}.

\begin{figure}[ht]
	\includegraphics[width=3.5in]{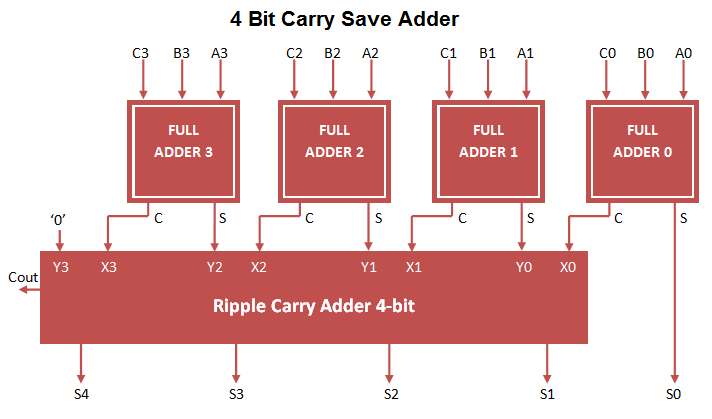}
    \caption[]{Carry save adder} 
\end{figure}

\section{Experimental Results and Output}

%


This section \cite{Pad47} shows the experimental results of the Denoising of the ECG signals of Conventional 5MR, 5MR as TMR (XOR), 5MR as TMR (XNOR), 5MR as Cascaded, 5MR with 4 to 1 Multiplexer in Xilinx Vivado EDA.In this technique, five identical ECG measurements are taken simultaneously, and the median value of these measurements is used as the final ECG signal. This approach can reduce the effect of noise and other artifacts on the ECG signal, and improve the accuracy of ECG analysis.

However, even with 5MR, ECG signals can still be affected by noise and artifacts, and therefore, ECG denoising techniques can be used in conjunction with 5MR to further improve the accuracy of ECG analysis. The ECG denoising of conventional 5MR is given below in the Fig 9.\cite{Roc47}

\begin{figure}[ht]
	\includegraphics[ width=3.5in]{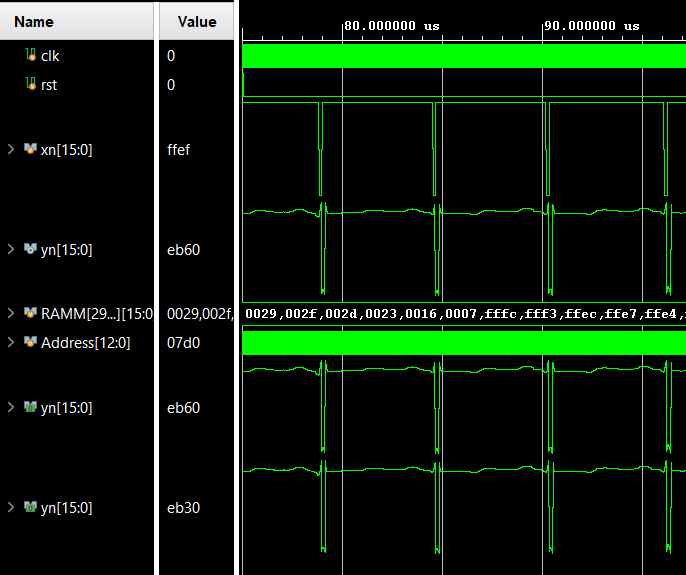}
    \caption[]{ECG Denoising of Conventional 5MR} 
\end{figure}

In the XOR-TMR configuration, three identical ECG measurements are taken simultaneously, and an XOR gate is used to combine the three measurements into a single output. If two of the three measurements are same, then XOR output is considered as the majority value. Otherwise, the output is the third measurement.

Combining 5MR with XOR-TMR can provide a higher level of redundancy to ECG systems, which can improve the reliability of ECG analysis. However, even with this level of redundancy, ECG signals can still be affected by noise and artifacts. Therefore, denoising techniques can be used in conjunction with 5MR and XOR-TMR configurations to further improve the accuracy of ECG analysis.
The ECG Denoising of 5MR Configuration with TMR(XOR as MUX) is given in Fig 10.

\begin{figure}[ht]
	\includegraphics[ width=3.5in]{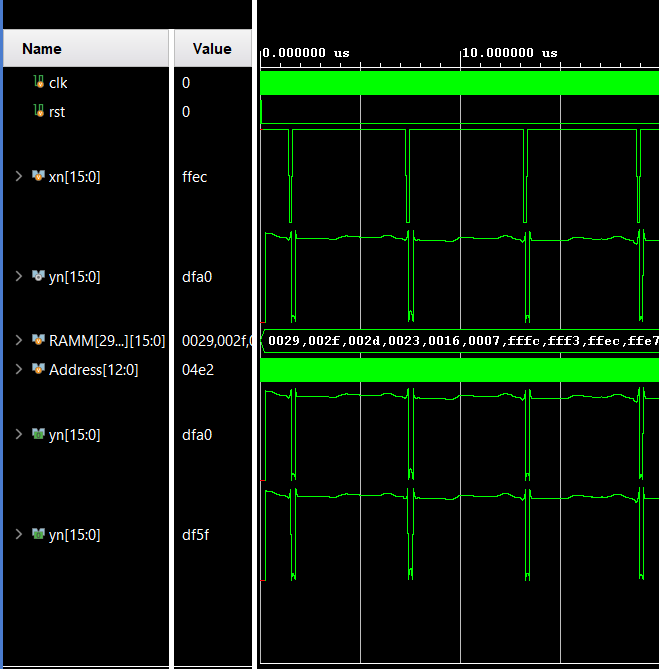}
    \caption[]{ECG Denosing on 5MR Configuration with TMR(XOR as MUX)} 
\end{figure}

In the XNOR-TMR configuration, three identical ECG measurements are taken simultaneously, and an XNOR gate is used to combine the three measurements into a single output. If all three measurements are the same, the XNOR output is that value. Combining 5MR with XNOR-TMR can provide a higher level of redundancy to ECG systems, which can improve the reliability of ECG analysis.

In summary, combining 5MR with XNOR-TMR can provide a higher level of redundancy to ECG systems, which can improve the reliability of ECG analysis. However, denoising techniques should also be used in conjunction with these configurations to further improve the accuracy of ECG analysis. The ECG denoising of 5MR Configuration with TMR(XNOR as MUX) is given below in the fig 11.

\begin{figure}[ht]
	\includegraphics[ width=3.5in]{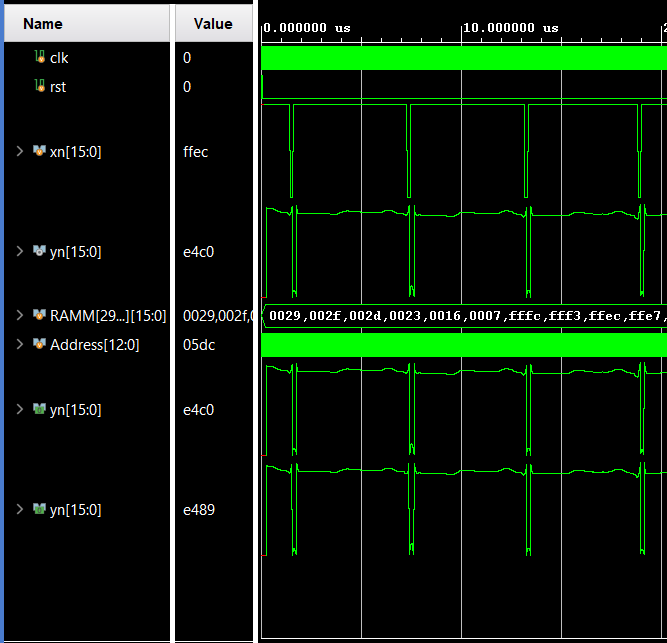}
    \caption[]{ECG Denosing on 5MR Configuration with TMR(XNOR as MUX)} 
\end{figure}

A cascaded TMR configuration can provide an even higher level of redundancy for the ECG system and improve the reliability of ECG analysis. However, like the other ECG configurations, Cascade TMR is not immune to noise and noise reduction techniques can be used to further improve the accuracy of the ECG analysis. Several denoising techniques can be used in conjunction with cascaded TMR configurations, including time-domain, frequency-domain, wavelet-based methods, and machine learning-based approaches. The ECG Denoising of 5MR as Cascaded TMR Configuration is given in Fig 12.\cite{Kar47}

\begin{figure}[ht]
	\includegraphics[ width=3.5in]{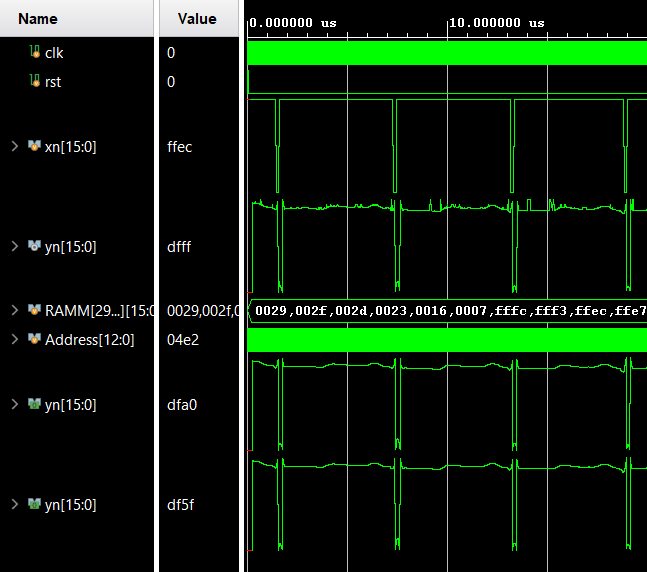}
    \caption[]{ECG Denosing of 5MR Cascaded with TMR} 
\end{figure}

In this configuration, four ECG measurements are taken simultaneously, and the MUX selects the most common value as the output. This approach is simpler and more efficient compared to traditional TMR implementations, which require three ECG measurements and a voting mechanism.

While using a MUX as a voting mechanism can reduce the hardware complexity of the system, it may not provide the same level of redundancy and reliability as traditional TMR or cascaded TMR configurations.\cite{Pat47} Using a 4 to 1 MUX as a voting mechanism can reduce the hardware complexity of ECG systems, but denoising techniques should still be used in conjunction with these configurations to improve the accuracy of ECG analysis.The ECG Denoising of 5MR with 4 to 1 MUX  is given in Fig 13.\cite{Maa47}

\begin{figure}[ht]
	\includegraphics[ width=3.5in]{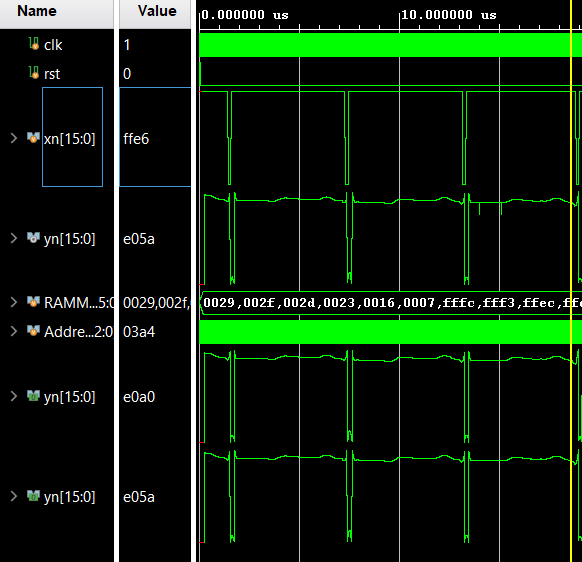}
    \caption[]{ECG Denosing of 5MR with 4 to 1 MUX} 
\end{figure}

All the proposed simulations and the implementation is done for Effective comparisons. In all the proposed FIR filters, ECG signal noise rejection is compared to other signals. Architectures reported  in the literature and  valid architectures is used in the EDA. The performance comparison is based on number of look-up tables (LUT's), slices and flip-flops. 

The value of the Table 1 shows comparison between Flipflops, LUT and Carry adder\cite{Ven47}. All architectures considered and  reported  are implemented in Xilinx EDA. This architecture is implemented in Xilinx xc7a100tcsg324-1 Family. The Below Fig 14 and Fig 15 are the bar graphs of Flip-fops and LUT.

\begin{figure}[ht]
	\includegraphics[ width=3.5in]{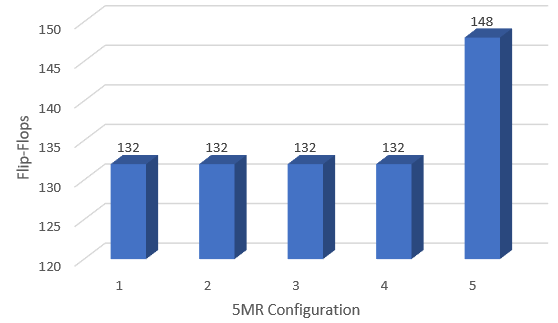}
    \caption[]{Flip-Flops Bar chart} 
\end{figure}

\begin{figure}[ht]
	\includegraphics[ width=3.5in]{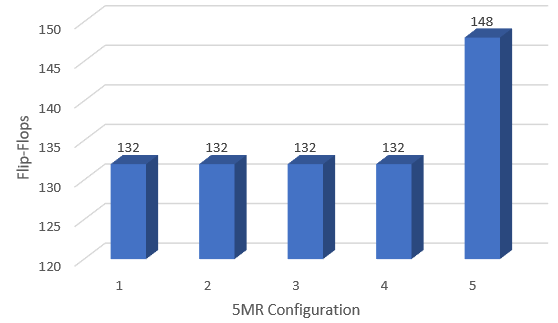}
    \caption[]{LUT Bar Chart} 
\end{figure}

After the synthesis of the Conventional 5MR there are 132 Flipflops that are used and 122 LUT that are used and 33 Carry logic is used\cite{Aga47}. The RTL Netlist view of Conventional 5MR Configuration is shown in Fig 16. In the synthesis of the 5MR Configuration with TMR(XOR as MUX) there are 132 Flipflops that are used and 122 LUT that are used and 33 Carry logic is used.

The synthesis of the 5MR Configuration with TMR(XNOR as MUX) there are 132 Flipflops that are used and 122 LUT that are used and 33 Carry logic is used.The synthesis of the 5MR Configuration as Cascaded TMR there are 132 Flipflops that are used and 225 LUT that are used and 57 Carry logic is used and the synthesis of the 5MR Configuration with with 4 to 1 MUX there are 148 Flipflops that are used and 244 LUT that are used and 57 Carry logic is used.

\begin{figure}[ht]
	\includegraphics[width=3.5in]{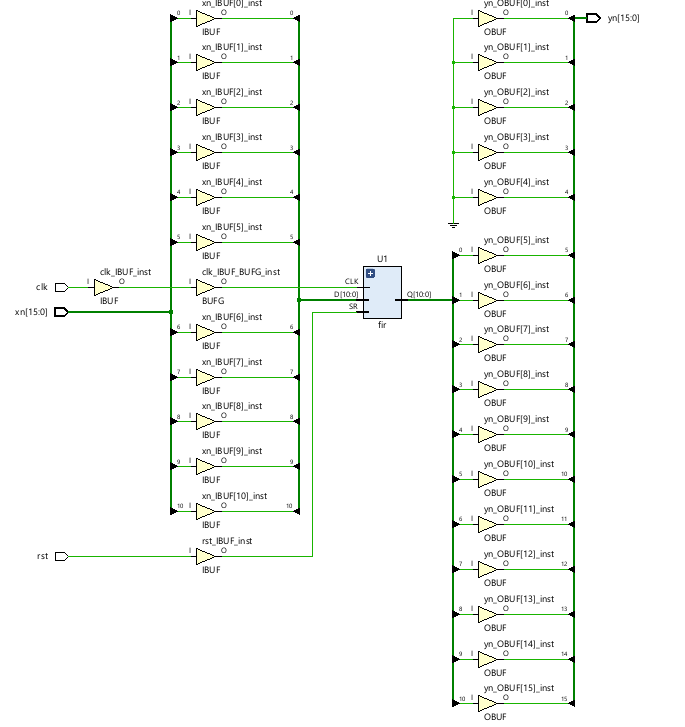}
    \caption[]{RTL Synthesized netlist view of Conventional 5MR Configuration} 
\end{figure}

\begin{figure}[ht]
	\includegraphics[width=3.5in]{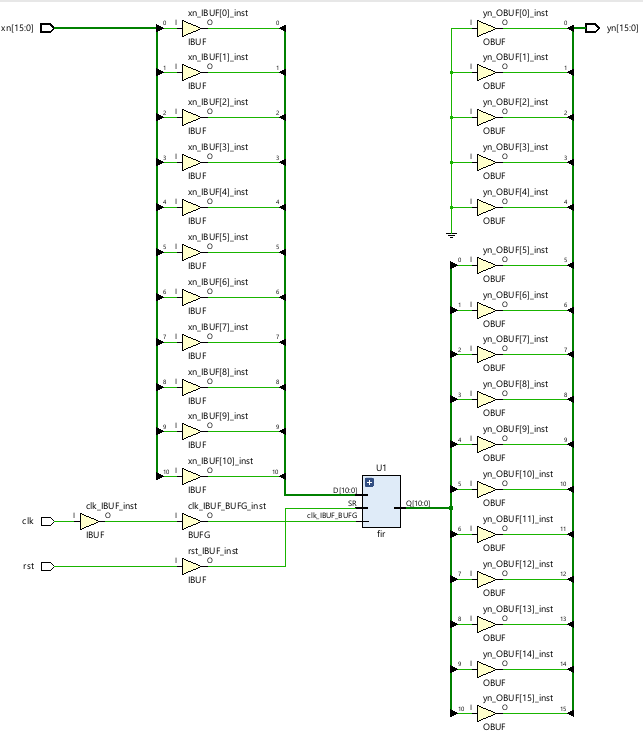}
    \caption[]{RTL Synthesized netlist view of 5MR Configuration with TMR(XOR as MUX} 
\end{figure}

\begin{figure}[ht]
	\includegraphics[width=3.5in]{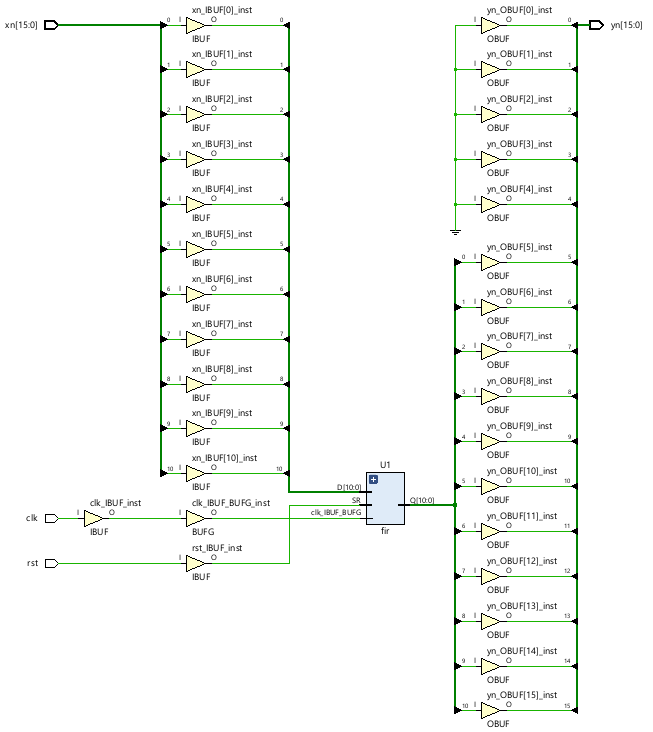}
    \caption[]{RTL Synthesized netlist view of 5MR Configuration with TMR(XNOR as MUX} 
\end{figure}

\begin{figure}[ht]
	\includegraphics[width=3.5in]{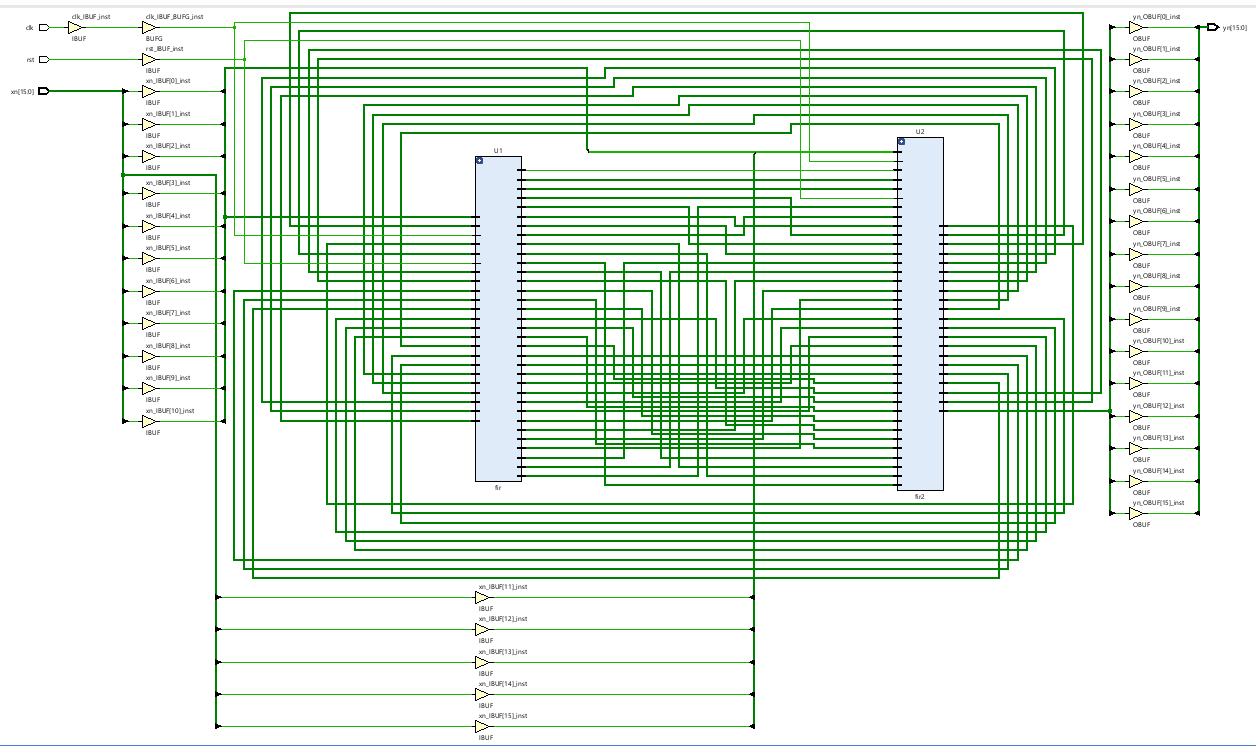}
    \caption[]{RTL Synthesized netlist view of 5MR Configuration as Cascaded TMR} 
\end{figure}

\begin{figure}[ht]
	\includegraphics[width=3.5in]{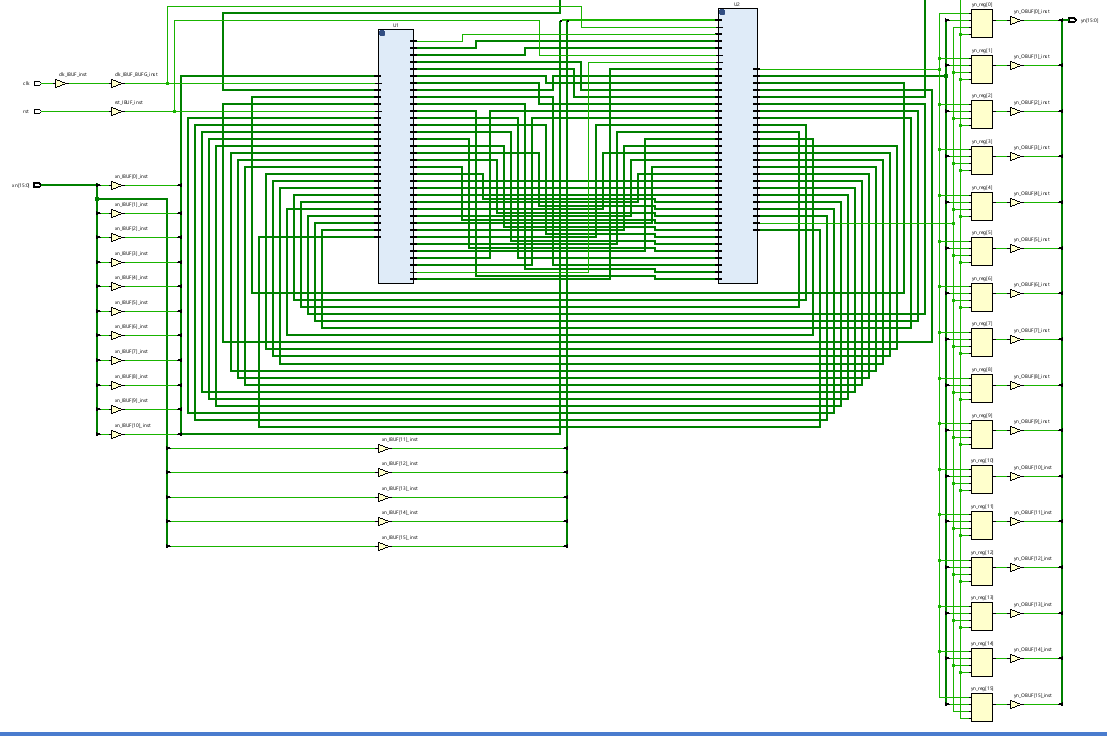}
    \caption[]{RTL Synthesized netlist view of 5MR with 4 to 1 MUX} 
\end{figure}

\begin{table}[ht] 
\centering 
\begin{tabular}{@{\small}llll@{}} \toprule
& {\footnotesize LUT} & {\footnotesize FF} & Carry \\ \midrule
Conventional 5MR Configuration & 122 & 132 & 33 \\

5MR Configuration with TMR(XOR) & 122 & 132 & 33 \\

5MR Configuration with TMR(XNOR) & 122 & 132 & 33 \\

5MR Configuration as Cascaded TMR & 225 & 132 & 57 \\  

5MR Configuration with 4 to 1 MUX & 244 & 148 &  57  \\ \bottomrule

\end{tabular} \caption{Comparison between Flip-Flops,LUT and Carry logic between the 5MR Configuration.} 
\end{table}

\clearpage

\newpage

\section{Conclusions}

The Fault-Tolerant Digital filters using 5MR configurations uses FIR filters with the approaches of Conventional 5MR Configuration, 5MR Configuration with TMR(XOR as MUX), 5MR Configuration with TMR(XNOR as MUX), 5MR as Cascaded with TMR, 5MR as with 4 to 1 MUX Configuration.
 Proposed architecture
Vedic Multiplier  is
high speed design to achieve minimal latency. Even in the proposed FIR
architecture, vedic multiplier structure
and carry-save adder-based implementation
consumes relatively little power and space
comparison with other reported FIR filter structures described in the literature. After simulation the results shows that all the 5MR Configurations clears the noise of the ECG signals in the Xilinx EDA tool and consumes less area.

\section{Acknowledgment}
With the deep sense of gratitude and immense Pleasure we like to thank Vellore Institute of Technology, Chennai for providing facilities with great infrastructures and many resources to complete this work.

\bibliography{main.bib}

\end{document}